# An Approach to 2D Signals Recovering in Compressive Sensing Context

Srdjan Stanković and Irena Orović

*University of Montenegro, Faculty of Electrical Engineering, 81000 Podgorica, Montenegro*
*Tel.: +382 67 516 975*
*Fax: +382 (20) 245 873*
*irenao@ac.me*

*Abstract-* **In this paper we study the compressive sensing effects on 2D signals exhibiting sparsity in 2D DFT domain. A simple algorithm for reconstruction of randomly under-sampled data is proposed. It is based on the analytically determined threshold that precisely separates signal and non-signal components in the 2D DFT domain. The algorithm operates fast in a single iteration providing the accurate signal reconstruction. In the situations that are not comprised by the analytic derivation and constrains, the algorithm is still efficient and need just a couple of iterations. The proposed solution shows promising results in ISAR imaging (simulated data are used), where the reconstruction is achieved even in the case when less than 10% of data is available.**

*Index Terms—Compressive sensing, signal reconstruction, missing data, ISAR imaging, sparsity*

## I. INTRODUCTION

Compressive sensing sensing (CS) was introduced as a new approach to data acquisition, that overcomes the common sampling approach [1]-[3]. In many applications traditional signal acquisition is resource-demanding and time-consuming. In the case of CS, only the small amount of samples is acquired in a random manner, and the signal is reconstructed using optimization algorithms. The CS assumes that the signal of interest is sparse, meaning that it can be represented by few nonzero coefficients in a suitable transform basis. The other requirement is related to the incoherent measurements [3]-[7]. Since the CS deals with the undersampled signals, the system of linear equations describing the relationship between the measurements and transform domain, is underdetermined and has infinitely many solutions [1],[2]. The original vector is reconstructed by searching for the sparsest solution that is consistent with the linear measurements. The commonly used approaches are convex relaxation based on $\ell_1$-norm minimization and greedy algorithms.

In this letter we focus on a simple and fast algorithm for the reconstruction of sparse 2D signals, which belongs to the group of greedy algorithm. The proposed solution is driven by the extensive analysis and derivations of missing samples phenomena in the 2D DFT domain. Namely, it has been shown that the main difficulty in signal reconstruction is spectral dispersion that appears as a consequence of vast majority of missing samples [5]. If we are able to estimate the spectral noise parameters induced by CS, we can distinguish between signal and non-signal/noise components [8],[9], which is primarily the aim of this work. Thus, the proposed solution assumes that the signal is sparse but does not use any a priori information about the expected number of components. In most applications the proposed algorithm will provide a high precision reconstruction in a single step. For more complex cases when the algorithm requirements are not meet, the reconstruction is done just a couple of iterations. This work is motivated by the inverse synthetic aperture radar (ISAR) images of a target, obtained using the 2D DFT of the received signal. Such a signal is usually sparse and consists of several pulses at the range and cross-range positions produced by reflecting points [10],[11]. According to the popular CS approaches, radar image can be perfectly reconstructed using far fewer samples than it was done so far (only 9% of samples are used in the experiments).

II. MISSING DATA MODELING IN THE CASE OF COMPRESSIVE SENSED 2D SIGNALS

*A. Spectral noise in the 2D DFT induced by the incomplete set signals*

Consider a 2D multicomponent signal in the form:

$$s(x,y) = \sum_{i=1}^{K} A_i e^{j2\pi k_{x_i} x/N_x} e^{j2\pi k_{y_i} y/N_y} \quad (1)$$

where $A_i$ denotes the *i*-th component amplitude, while $kx_i$ and $ky_i$ denote frequencies of the *i*-th component. The 2D DFT of such a signal can be written as:

$$S(k_x,k_y) = N_x N_y \cdot \sum_{x=1}^{N_x}\sum_{y=1}^{N_y}\sum_{i=1}^{K} A_i e^{-j2\pi(k_x-k_{x_i})x/N_x} e^{-j2\pi(k_y-k_{y_i})y/N_y} \quad (2)$$

Furthermore, we observe the following (full) set of samples ($N_x$ samples in *x* direction, and $N_y$ along the *y* dimension):

$$\mathbf{Z} = \{z(x,y): x \in \{1,...,N_x\},\ y \in \{1,...,N_y\}\}, \quad (3)$$

where $\quad z(x,y) = A_i e^{-j2\pi(p_i x/N_x + q_i y/N_y)},\ p_i = k_x - k_{x_i},\ q_i = k_y - k_{y_i}.$

In the case of compressed sensed signal, instead of the full set of samples **Z**, we deal with the partial set **W**:

$$\mathbf{W} = \{w(x,y): x \in \{1,...,M_x\},\ y \in \{1,...,M_y\}\} \subset \mathbf{Z}, \quad (4)$$

having $M_x$ samples available along *x* direction and $M_y$ samples available along *y* direction. Obviously, we need to deal with missing samples and consequently, an appropriate model of the missing samples occurrence is required [5]. In that sense, consider the missing samples as zero values occurring due to an additive noise influence given in the form:

$$\varepsilon(x,y) = \begin{cases} -z(x,y) = -e^{-j2\pi px/N_x}e^{-j2\pi qy/N_y} \\ 0, \quad \text{for } (x,y) \text{ belonging to available samples} \end{cases}$$

The 2D DFT of the available samples **W** can be written as:

$$F(k_x,k_y) = \sum_{x=0}^{M_x-1}\sum_{y=0}^{M_y-1} z(x,y) + \varepsilon(x,y). \qquad (5)$$

In order to characterize the influence of missing samples, in the sequel we will observe the following cases.

a) If $k_x = k_{xi}$, $k_y = k_{yi}$ the expectation is given as:

$$\begin{aligned} E\{F(k_x,k_y)\} &= A_1 E\left\{\sum_{x=1}^{M_x}\sum_{y=1}^{M_y} e^{-j2\pi[(k_x-k_{x_1})x/N_x+(k_y-k_{y_1})y/N_y]}\right\} \\ &\quad \ldots + A_K E\left\{\sum_{x=1}^{M_x}\sum_{y=1}^{M_y} e^{-j2\pi[(k_x-k_{x_K})x/N_x+(k_y-k_{y_K})y/N_y]}\right\} \end{aligned} \qquad (6)$$

which equals to:

$$E\{F(k_x=k_{xi},k_y=k_{yi})\} = A_i M_x M_y.$$

b) If $k_x \neq k_{xi} \vee k_y \neq k_{yi}$, the expectation is $E\{F(k_x,k_y)\} = 0$, since:

$$E\{\sum_{x=1}^{M_x} \exp(-j\frac{2\pi}{N_x}(k_x-k_{x_i})x)\} = 0 \text{ for } k_x \neq k_{xi}$$

(which also holds for the coordinate $y$ and $k_y \neq k_{yi}$). Now, we can determine the variance of noise introduced in $F(k_x, k_y)$:

$$\begin{aligned} \sigma_i^2\{F(k_x \neq k_{xi}, k_y \neq k_{yi})\} &= M_x M_y \cdot E\{z(x,y)z^*(x,y)\} + M_x M_y (M_x M_y - 1) \cdot E\{z(x,y)z^*(m,n)\} = \\ &= A_i M_x M_y + A_i M_x M_y (M_x M_y - 1)\frac{-1}{N_x N_y - 1} \end{aligned}$$

Finally, the total variance that appears in the 2D DFT as a reflection of missing samples influence, can be calculated according to (for *K* components):

$$\sigma^2 = \sum_{i=1}^{K} A_i M_x M_y \frac{N_x N_y - M_x M_y}{N_x N_y - 1}. \tag{7}$$

Here, it is important to emphasize that the same analysis holds for radar signal model, when continuous wave radar transmits a signal in the form of series of chirps and the missing samples appear in one or different chirps. In analogy, the received signal will consist of *K* components reflected from different points. After the distance compensation, the received signal can be written as in (1):

$$R(x, y) = \sum_{i=1}^{K} r_i e^{j2\pi \alpha_x x / N_x} e^{j2\pi \beta_y y / N_y}, \tag{8}$$

where the $A_i = r_i$ is the *i*-th reflection coefficient of the target, while $k_x = \alpha_x$ and $k_y = \beta_y$ correspond to the parameters proportional to the velocity and range, respectively.

III. SIMPLE AND FAST ALGORITHM FOR RECONSTRUCTION OF 2D SIGNALS – SFAR-2D

The estimated noise parameters such as noise variance, can be efficiently used to distinguish between signal components, i.e., useful information, and noise components. Note that the noise components originate from missing samples, while the noise variance directly depends on the missing samples number. Also, the noise may appear from an external source, making the extraction of signal components even more difficult. On the other side, if we are able to estimate level of noise and to detect positions of useful signal components above the noise, then it would be possible to reconstruct the original signal, as it is shown in the sequel.

Consider the noise model introduced in Section II. According to the central limit theorem, the real and imaginary parts of the 2D DFT values at the position where only noise exists (none of the $K$ signal components occur) can be described by the Gaussian distribution. The corresponding probability density function (pdf) for the absolute values of 2D DFT is modeled by the Rayleigh function. The probability that all DFT values at noise positions are lower than the certain threshold $\chi$ is given by:

$$P(\chi) = \left(1 - \int_{\chi}^{\infty} \frac{2p}{\sigma^2} e^{-\frac{p^2}{\sigma^2}} dp \right)^{N_x N_y - K} \approx \left(1 - \exp(-\frac{\chi^2}{\sigma^2})\right)^{N_x N_y}. \qquad (9)$$

where $N_x N_y$ is the total number of samples in 2D DFT of size $N_x \times N_y$, while $K$ is the number of signal components and $K \ll N_x N_y$. For instance, we can set a fixed probability of error $P_{FIX}=0.99$ and calculate the threshold as follows:

$$\chi = \sigma \sqrt{-\log(1 - P_{FIX}^{\frac{1}{N_x N_y}})}. \qquad (10)$$

It means that with the probability $P_{FIX}$ all noise components will be below the signal components. Therefore, if we observe the set of positions representing the support of available samples within the 2D signal:

$$(x, y) \in \Omega = \{(x_1, y_1), (x_2, y_2), ..., (x_{M_x}, y_{M_y})\}, \qquad (11)$$

then the vector of measurements containing only $M_x M_y$ out of $N_x N_y$ samples of original signal can be written in the form:

$$\mathbf{y} = s(\Omega). \qquad (12)$$

Furthermore, the Fourier transform of the full signal set is denoted using the 2D DFT matrix $\mathbf{\Psi}$ ($N_x N_y \times N_x N_y$), obtained as Kronecker product of two DFT matrices:

$$\Psi = \mathbf{DFT}_{N_x \times N_y} \otimes \mathbf{DFT}_{N_x \times N_y}.$$

Consequently, the initial Fourier transform of available measurements can be written as follows:

$$\mathbf{Y} = \Psi(\Omega)' \mathbf{y} \qquad (13)$$

where $\mathbf{y}$ is of size ($M_x M_y \times 1$), while $\Psi(\Omega)$ contains only $M_x M_y$ rows from $\Psi$ that correspond to available time instants $\Omega$. Vector $\mathbf{Y}$ contains the 2D DFT coefficients of the incomplete set of measurements $\mathbf{y}$, and thus contains the noise due to the missing samples. In order to select the signal components among the noise, we can apply threshold $\chi$ to obtain the frequency support of the signal:

$$\vec{\mathbf{k}}(k_x, k_y) = \arg\{|\mathbf{Y}| > \chi\}. \qquad (14)$$

Finally, we can obtain the reconstructed signal by solving the following minimization problem:

$$\mathbf{S} = \left(\Psi(\Omega, \vec{\mathbf{k}})' \Psi(\Omega, \vec{\mathbf{k}})\right)^{-1} \Psi(\Omega, \vec{\mathbf{k}})' \mathbf{y} \qquad (15)$$

where $\mathbf{S}$ contains only the recovered DFT values of signal components, while $\Psi(\Omega, \vec{\mathbf{k}})$ is a matrix that contains columns defined by selected frequency set $\vec{\mathbf{k}}$ and rows defined by the positions of measurements $\Omega$. Based on the previous discussion, we can define a simple and fast algorithm that can be easily applied to the sparse 2D signals in the 2D DFT domain. Fig. 1 summarizes the proposed SFAR-2D algorithm in the pseudo-code format.

---

**Input** $i=1$, $y$, $\Omega$, $\mathbf{p}=\varnothing$,
**Output:**
1: **Calculate** $\sigma^2 = \sum_{i=1}^{K} A_i M_x M_y (N_x N_y - M_x M_y)/(N_x N_y - 1)$
2: **Calculate** $\chi$ for a given $P_{FIX}=0.99$.
3: **Calculate** the initial DFT vector $\mathbf{Y}_i$ that corresponds to $y$
4: $\vec{\mathbf{k}} = \arg\{|\mathbf{Y}|>\chi\}$, $\text{Card}\{\mathbf{k}\} \leq M$
5: $\mathbf{S} = \left(\Psi(\Omega,\vec{\mathbf{k}})'\Psi(\Omega,\vec{\mathbf{k}})\right)^{-1} \Psi(\Omega,\vec{\mathbf{k}})'\mathbf{y}$

---

**Fig. 1. Pseudo code describing SFAR-2D algorithm – single step version**

In the case of large number of missing samples, some DFT components will be masked by the derived spectral noise. Therefore, we need slightly modified form, which is summarized in Fig. 2. This is actually an iterative form of the previous algorithm. Namely, in order to reveal weaker components it is important to remove the contribution of stronger ones. Thus, in each iteration $i$, a set of components on positions $\mathbf{k}_i$ is detected and removed. After just a couple of iterations, it is possible to reconstruct all signal components.

---

**Input** $i=1$, $y$, $\Omega$, $\mathbf{p}=\varnothing$, $\Psi=\mathbf{DFT}_{N_x \times N_y} \otimes \mathbf{DFT}_{N_x \times N_y}$, $\Gamma=\Psi^{-1}$
**Output:**
1: **Calculate** $\sigma^2 = \sum_{i=1}^{K} A_i M_x M_y (N_x N_y - M_x M_y)/(N_x N_y - 1)$
2: **Calculate** $\chi$ for a given $P_{FIX}=0.99$.
3: **Calculate** the initial DFT vector $\mathbf{Y}_i$ that corresponds to $y$
4: **Repeat** $\vec{\mathbf{k}} = \arg\{|\mathbf{Y}|>\chi\}$, $\text{Card}\{\mathbf{k}\} \leq M$
5:   **Solve:** $\mathbf{S} = \left(\Psi(\Omega,\vec{\mathbf{k}})'\Psi(\Omega,\vec{\mathbf{k}})\right)^{-1} \Psi(\Omega,\vec{\mathbf{k}})'\mathbf{y}$
6:   **Update** $y$ by subtracting contribution of $\mathbf{Y}(\vec{\mathbf{k}})$:
7:     $\mathbf{y} = \mathbf{y} - \Gamma(\Omega,\vec{\mathbf{k}})\mathbf{Y}(\vec{\mathbf{k}})$
8:   **Update Y** for new vector $y$: $\mathbf{Y} = \Psi(\Omega)'\mathbf{y}$
9:   **Update** $A^2 = \sum|\mathbf{y}|^2/M_x M_y$ and $\sigma^2$.
10: **until** all components are detected.

---

**Fig. 2. Pseudo code describing SFAR-2D algorithm – iterative version**

*External noise presence:* If the observed compressive sensed signal is corrupted by external noise, this influence needs to be included in the algorithm as well. Namely, assuming that the variance of external noise is $\sigma^2_\varepsilon$, the step 1 of both algorithm versions need to include this parameters as follows:

$$\sigma_{new}^2 = \sigma_\varepsilon^2 + \sigma^2.$$

IV. EXAMPLES

*Example1:* Consider a signal that consists of 12 reflecting points:

$$R(x,y) = \sum_{i=1}^{K} r_i e^{j2\pi\alpha_x x/N_x} e^{j2\pi\beta_y y/N_y},$$

with reflecting coefficients varying in the range $2<r_i<3$. Only 9% of the total number of samples are available (91% of samples are missing). In order to illustrate the desired and expected result of the reconstruction procedure, the 2D DFT of the original signal (ISAR image) with full set of samples is shown in Fig. 3 (a and b).

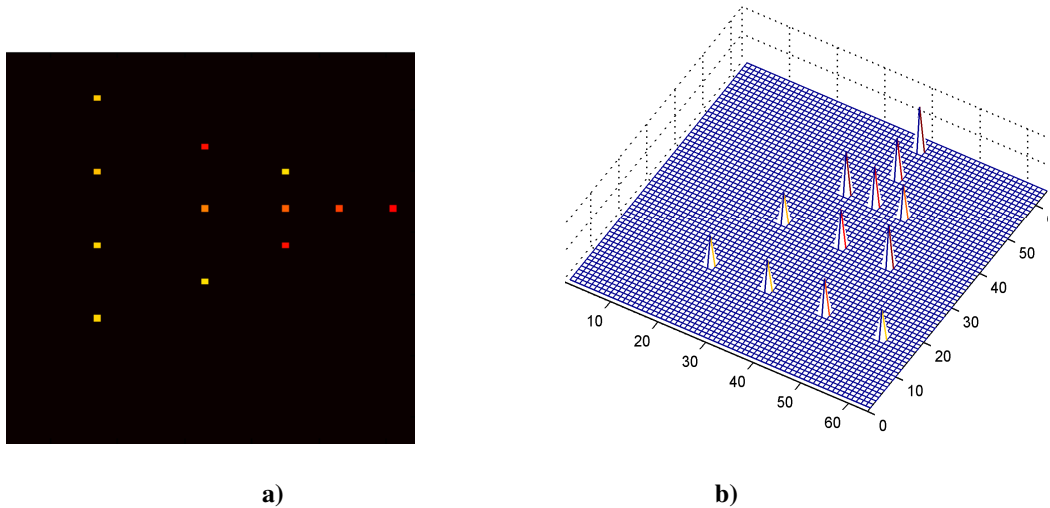

a)  b)

**Fig. 3. 2DDFT of considered radar signal, a) 2D view, b) 3D view**

Apart from the noise that is produced in the 2D DFT as a consequence of a large number of missing samples, the signal is also corrupted by the external noise. The signal reconstruction procedure is done according to the proposed **SFAR-2D algorithm – single step version**. Fig. 4.a illustrates the result of applying the threshold $\chi$, which detects all 12 components from the 2D DFT obtained using only 9% of available samples. Based on the detected parameters of components, the signal is successfully reconstructed. The final result of reconstruction is shown in Fig. 4.b.

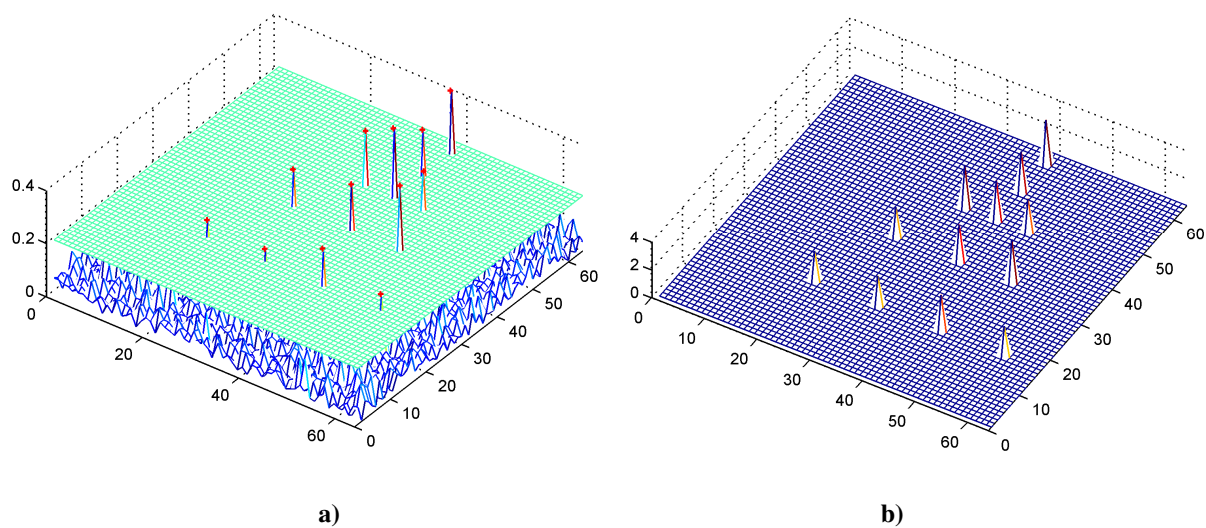

a)               b)

**Fig. 4. Result of applying the proposed algorithms: a) Components of 2D signal detected all in a single step, b) Recovered 2D DFT of the considered signal**

*Example2:* In this example, we consider the case when it is not possible to detect all components at once. The signal also consists of 12 components, where some of the components are significantly smaller than the others. The energy of components will be lower compared to the variance of noise that appear in the 2D DFT as a reflection of missing samples influence. Also, we deal with more than 90% of missing samples. In this case, for the full signal reconstruction, it is necessary to apply the iterative version of the algorithm, since it is not possible to detect all components in a single step.

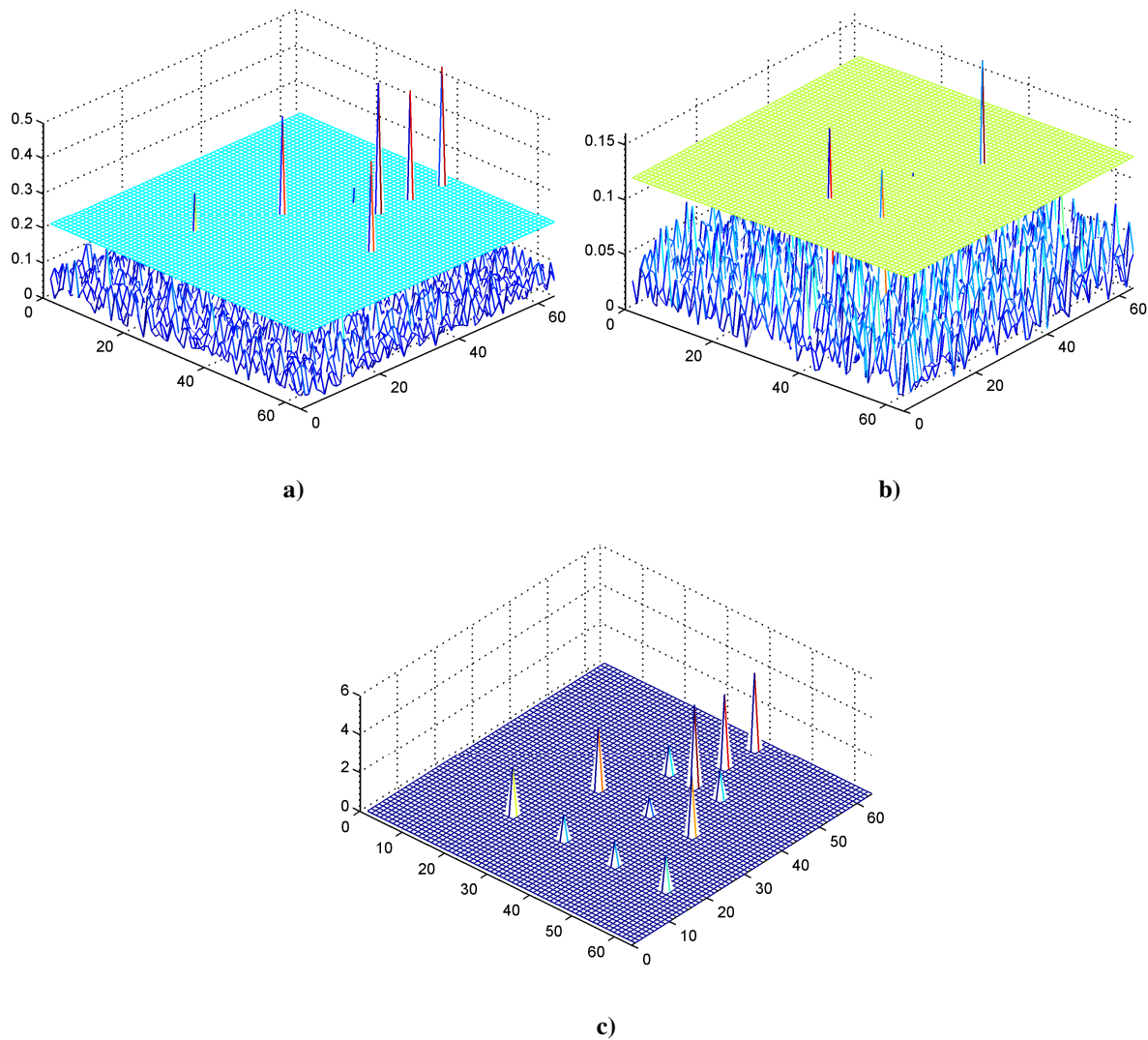

**Fig. 5. a) Components of 2D signal detected in the first step (other components are still masked by noise), b) Components of 2D signal detected in the second step, c) Recovered 2D DFT**

After detecting the first set of components, we eliminate their influence within the spectral representation and continue with procedure using new signal parameters (according to Fig. 2). The results of applying the threshold are presented in Fig. 5a (first iteration), and Fig. 5b (second iteration). It is important to emphasize that the signal is completely recovered after just couple of iterations (Fig 5.c), which is an advantage over other iterative procedure used for CS reconstruction.

V. Conclusion

In this paper, the analytical expression is derived to describe the spectral dispersion that appears in the 2D DFT domain as a consequence of missing data. The analysis is motivated by the ISAR applications where 2D DFT is assumed as a sparsity domain. The proposed analysis reveals an important feature: the number of missing samples can be used to define the threshold which separates signal and non-signals components, and facilitates CS signal reconstruction. Two versions of the algorithm are proposed: the first one is based on the simple single step solution, while the second one performs additional iteration/s when some components are much weaker than the others.

Acknowledgement

This work is supported by the Montenegrin Ministry of Science, project grant funded by the World Bank loan: CS-ICT "New ICT Compressive sensing based trends applied to: multimedia, biomedicine and communications".